# A NEW APPLICATION OF THE LUNAR LASER RETROREFLECTORS: SEARCHING FOR THE "LOCAL" HUBBLE EXPANSION

Yu. V. Dumin

*IZMIRAN, Russian Academy of Sciences, Troitsk, Moscow reg., 142190 Russia*

## ABSTRACT

Precise measurements of the Earth–Moon distance by the lunar laser ranging (LLR), which begun in the early 1970's, contributed significantly to geodesy, geophysics, and lunar planetology, as well as enabled astrophysicists to perform several fine tests of the relativistic gravitational field theory (General Relativity). Yet another promising application of LLR arises just now in the context of recent cosmological models, whose dynamics is substantially affected by some kinds of the dark matter (or the so-called "dark energy") uniformly distributed in space, and should therefore be accompanied by some residual Hubble expansion at any spatial scales, particularly, in the Earth–Moon system. The "local" Hubble expansion can be revealed by comparing the rate of increase in the lunar semi-major axis measured by LLR (which should be produced both by the well-known tidal exchange of angular momentum between the Earth and Moon and the local Hubble expansion) with the same quantity derived indirectly from astrometric data on the Earth's rotation deceleration (which is produced only by the tidal interaction). Such analysis really points to the discrepancy 1.3 cm/yr, which corresponds to the local Hubble constant $H_0^{(\rm loc)} = 33 \pm 5$ (km/s)/Mpc. This value is about two times less than at intergalactic scales but many orders of magnitude greater than was predicted in earlier theoretical works.

## INTRODUCTION

Precise measurements of the Earth–Moon distance by using the ultrashort laser pulses—the lunar laser ranging (LLR)—are carried out for over 30 years, after the installation of several retroreflectors on the lunar surface in the course of *Apollo* (USA) and *Lunakhod* (USSR, in collaboration with France) space missions (e.g. reviews by Dickey et al., 1994; Nordtvedt, 1999; Samain et al., 1998). The typical accuracy of these measurements was:
- ~25 cm in the early 1970's,
- 2–3 cm in the late 1980's, and
- a few millimeters at the present time.

LLR works contributed significantly to astrometry, geodesy, geophysics, lunar planetology, and gravitational physics (General Relativity). Among the most important results related to the last-mentioned subject are:
- verification of the Strong Equivalence Principle with accuracy up to $\sim 10^{-13}$,
- determination of the first post-Newtonian parameters in the gravitational field equations,
- detection of the so-called geodetic precession of the lunar orbit, and
- imposing the observational constraints on time variations in the gravitational constant $(dG/dt)/G$ with accuracy $10^{-11}$ per year.

The aim of the present report is to draw attention to yet another promising application of LLR in General Relativity and cosmology—measuring the rate of local Hubble expansion,—which is especially interesting in the context of recent "dark-energy-dominated" cosmological models.

## SECULAR VARIATIONS IN THE EARTH–MOON DISTANCE

Despite the considerable advances listed in the Introduction, there is a long-standing unresolved problem in the interpretation of LLR data—anomalous secular increase in the lunar semi-major axis (e.g. Pertsev, 2000). In general, such increase is well-known and can be **partially** explained by tidal interaction between the Earth and Moon (e.g. Kaula, 1968). For the sake of clarity, let us briefly remind the basic features of this effect.

**The Mechanism of Tidal Interaction between the Earth and Moon**

As is shown in Figure 1, a gravitational action by the Moon produces a tidal bulge on the Earth's surface. Because of the relaxation processes in the Earth's interiors and oceans, this bulge is not perfectly symmetric about the Earth–Moon line but slightly shifted in the direction of Earth's rotation.[1] As a result, there is a torque moment, which decelerates a proper rotation of the Earth and accelerates an orbital motion of the Moon; so that the mean Earth–Moon distance increases.

From the angular momentum conservation law

$$I_E \frac{d}{dt}\Omega_E + m_M \frac{d}{dt}(R^2 \Omega_{ME}) = 0 \quad (1)$$

and the relation between the lunar orbital velocity and its distance from the Earth

$$\Omega_{ME} = G^{1/2} m_E^{1/2} R^{-3/2}, \quad (2)$$

it can be easily found that the rate of increase in the lunar semi-major axis $dR/dt$ is related to the rate of deceleration of the Earth's rotation $dT_E/dt$ by the well-known formula

$$dR/dt = k\, dT_E/dt, \quad (3)$$

where

$$k = 4\pi\, G^{-1/2} I_E m_E^{-1/2} m_M^{-1} R^{1/2} T_E^{-2} = 1.81 \cdot 10^5 \text{ cm/s}. \quad (4)$$

Here, $\Omega_E$ and $T_E$ are the angular velocity and period of the proper rotation of the Earth, $\Omega_{ME}$ is the angular velocity of orbital rotation of the Moon about the Earth, $R$ is the distance between them, $I_E$ is the Earth's moment of inertia, $m_E$ and $m_M$ are the terrestrial and lunar masses, and $G$ is the gravitational constant. (Within the accuracy required here, we can neglect the ellipticity of the lunar orbit.)

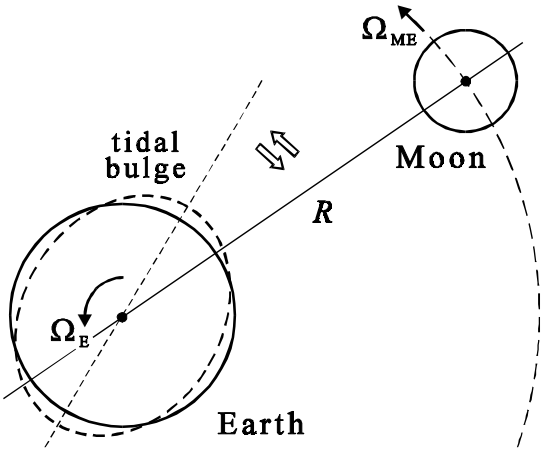

Fig. 1. Tidal exchange of angular momentum between the Earth and Moon.

So, if $dT_E/dt$ is known from independent measurements of the Earth's rotation deceleration with respect to the distant objects, then Eqs. (3)–(4) enable us to calculate $dR/dt$. Next, by comparing the predicted value of $dR/dt$ with the one actually measured by LLR, we can conclude if there are some influences on the secular increase in the Earth–Moon distance apart from the tidal effects.

**LLR versus the Angular Astrometric Measurements**

Astrometric measurements of the Earth's rotation are carried out from approximately 1650 till now (e.g. Chapter 1 in book by Sidorenkov, 2002); and they revealed a wide range of variations in the Earth's angular velocity with periods from a few hours to dozens of years, as well as a secular trend (systematic deceleration). There is a number of geophysical effects responsible for these variations, such as the lunar and solar tides, seasonal redistribution of air mass in the Earth's atmosphere, exchange of angular momentum between the mantle and core, etc. While the short-period variations were identified and interpreted quite reliably, the data on the variations with the longest periods (60–70 yr) are somewhat contradictory because of insufficient length of the time series and low accuracy of early observations (1650–1850), as illustrated, for example, in Figure 1.1 by Sidorenkov (2002).[2]

---

[1] To avoid misunderstanding, it should be emphasized that the field of gravitational perturbations produced on the Earth by the lunar attraction is, of course, perfectly aligned with the Earth–Moon direction; but the resulting mass distribution in the deformed Earth depends also on boundary conditions for the velocity, accounting for the rigid-body Earth's rotation. Since the angular velocity of proper rotation of the Earth is greater than the velocity of orbital rotation of the Moon, the tidal bulge turns out to be shifted in the direction of the Earth's rotation.

Despite the above-mentioned uncertainties, a secular deceleration of the Earth's rotation derived from the various sets of telescopic observations turns out to be approximately the same and equals $(dT_E/dt)^{(tel)} = 1.4 \cdot 10^{-5}$ s/yr (Pertsev, 2000). So, according to Eqs. (3)–(4), the predicted rate of secular evolution of the lunar orbit should be $(dR/dt)^{(tel)} = 2.53$ cm/yr. On the other hand, immediate measurements of the Earth–Moon distance by LLR technique give the appreciably greater value $(dR/dt)^{(LLR)} = 3.82$ cm/yr (Dickey et al., 1994). A presence of the discrepancy $\Delta(dR/dt) = (dR/dt)^{(LLR)} - (dR/dt)^{(tel)} = 1.29$ cm/yr, in principle, was well recognized in the last decade but still not explained satisfactorily by any particular geophysical phenomenon.

From our point of view, a promising interpretation of the anomaly $\Delta(dR/dt)$ may be based on the cosmological Hubble expansion in the local space environment, which should contribute to $(dR/dt)^{(LLR)}$ but will not manifest itself in $(dR/dt)^{(tel)}$. As follows from the standard relation

$$\Delta(dR/dt) = H_0^{(loc)} R , \qquad (5)$$

the local Hubble constant should be

$$H_0^{(loc)} = 33 \pm 5 \text{ (km/s)/Mpc} . \qquad (6)$$

Let us emphasize that only statistical errors were taken into account in the above-written expression. Besides, there may be systematic errors, related to the geophysical assumptions used in deriving Eqs. (3)–(4). The most important of them is ignoring a probable secular variation in the Earth's moment of inertia $I_E$.

**COSMOLOGICAL INFLUENCES ON THE DYNAMICS OF THE EARTH–MOON SYSTEM**

At first sight, the Hubble constant given by Eq. (6) looks very surprising: on the one hand, it is about two times less than the commonly-accepted value at intergalactic scales; but on the other hand, it is many orders of magnitude greater than some purely theoretical predictions of the Hubble expansion in small gravitationally-bound systems. Let us discuss this subject in more detail.

**Theoretical Estimates of the Local Hubble Expansion**

In the most of textbooks on General Relativity (e.g. Misner et al., 1973, §27.5), it is postulated *a priori* that sufficiently small gravitationally-bound systems (such as the planetary systems or galaxies) do not experience the universal Hubble expansion. Besides, there is a number of papers whose authors tried to justify the above statement by rigorous mathematical consideration (see, for example, the recent work by Domínguez and Gaite, 2001, and references therein). Unfortunately, most of these studies were based on the oversimplified interpretation of Hubble expansion in a small volume as the Newtonian gravitational action of distant matter on the local test masses. The resulting effect was found to be extremely small or absent at all.[3]

From our point of view, the above-mentioned approach is not satisfactory, because the equations of Newtonian gravity can be rigorously derived by linearization of the Einstein equations of General Relativity only against the background of flat (Minkowski) space–time. On the other hand, **a self-consistent calculation of the local Hubble expansion requires linearization against the background of Friedmann–Robertson–Walker (FRW) metric** (which by itself represents a nonvacuum solution of the Einstein equations). Unfortunately, accurate mathematical treatment of the respective problem is very difficult, especially when the average FRW metric is produced by a matter distribution strongly irregular at small scales.

---

[2] Fortunately, since there are no clearly-expressed periods larger than 60–70 yr, one can expect that all poorly-known periodic variations will be averaged out over the sufficiently long interval under consideration (300–350 yr) and, therefore, will not affect the value of systematic trend.

[3] Moreover, the effect calculated by Domínguez and Gaite (2001) can even change its sign in some cases (i.e. represents contraction instead of expansion). This is yet another evidence that the phenomenon under consideration is absolutely unrelated to the local Hubble effect.

In general, the effect of FRW background should not be ignored, as can be illustrated by the following example. Let us consider the idealized cosmological model whose dynamics is completely determined by Λ-term (or some other type of matter with perfectly-uniform spatial distribution); while the compact objects are only the tracers, whose energy–momentum tensor is infinitely small. Then, these objects will evidently experience Hubble expansion at small scales with the same amplitude as everywhere in the Universe.

Next, let us return to the more realistic cosmological models, widely discussed now, where about 50–70% of contribution to the energy–momentum tensor is produced by some kinds of an unclumped dark matter, or the so-called "dark energy" (such as Λ-term, "quintessence", inflaton-like scalar field, and so on), while the other 30–50% are caused by the matter experiencing a gravitational instability and forming the compact objects. Although, as was already mentioned above, we cannot estimate reliably the influence of the irregularly-distributed distant matter on the local background metric, it can be reasonably assumed that the effect of the unclumped dark matter still persists. **So, the local Hubble expansion should have the amplitude at least about one-half its value at intergalactic scales** (where it is formed both by the clumped and unclumped matter). **This is just the value given by Eq. (6).** (In principle, the distant matter concentrated in compact objects also cannot be excluded *a priori* as a probable source of the local FRW metric, and this subject requires a more careful theoretical study.)

**On the Effect of Unclumped Dark Matter at Small Intergalactic Scales**

At last, it is interesting to mention about yet another manifestation of the unclumped dark matter in the dynamics of Hubble expansion at the sufficiently small scales. The recent observations and computer simulation of the galactic velocities around the Milky Way (e.g. by Ekholm et al., 2001; and Horellou, 2002) revealed that a linear "quiescent" Hubble flow begins at the distances as short as ~1–2 Mpc. On the other hand, it was commonly believed before that a regular Hubble flow can be traced only from the scales ~5–10 Mpc, at which the distribution of visible matter can be averaged out and looks like a uniform background. Presence of the well-formed Hubble expansion at the substantially less scales was explained by Chernin et al. (2000) and Horellou (2002) by dominant contribution of the uniformly-distributed dark matter (or "dark energy") to the energy–momentum tensor of FRW model at small intergalactic distances.

**PROBLEMS OF INTERPRETATION AND PERSPECTIVES OF REFINEMENT**

Of course, a cosmological nature of the anomalous increase in the Earth–Moon distance is not confidently established by now. It may be also caused by some geophysical artefacts, such as a secular increase in the Earth's moment of inertia $I_E$ (which violates validity of the assumptions used in deriving Eqs. (3)–(4)) or quasi-periodic exchange of angular momentum between the mantle and core with periods comparable to the length of the available series of observations, 300–350 yr (which will be indistinguishable from the secular variation of the Earth's rotation). So, such effects need to be discriminated very carefully.

**On the Effect of Temporal Variations in the Gravitational Constant**

As is known, the temporal variations in the effective gravitational constant are predicted by some alternative theories of gravitation. So, such variations were actively sought during the last decades by various methods, particularly, the lunar laser ranging and microwave ranging of martian spacecraft. As was already mentioned in the Introduction, the upper limit on $(dG/dt)/G$ was found to be about $10^{-11}$ per year.

The secular increase in a planetary orbit caused by the changing gravitational constant is given by the relation

$$dR/dt = -(1/G)(dG/dt) R , \qquad (7)$$

which is formally equivalent to Eq. (5) for the local Hubble expansion if $H_0^{(\text{loc})}$ is identified with $-(1/G)(dG/dt)$. So, the above-mentioned value $10^{-11}$/yr can be recalculated to the upper limit on the rate of local Hubble expansion, $H_0^{(\text{loc})} \leq 10$ (km/s)/Mpc, which turns out to be substantially less than the value given by Eq. (6).

Unfortunately, such formal recalculation is meaningless from the physical point of view, because searching for the temporal variations in the gravitational constant is commonly carried out under assumption of conservation of orbital angular momentum of the Moon, while this condition evidently breaks down by Hubble expansion. So, the secular increase in the lunar semi-major axis produced by the local Hubble effect will mimic a tidal influence of the Earth rather than temporal variation of the gravitational constant.[4]

**Perspectives of Refinement**

One of the possible ways to verify the data on the anomalous increase in the lunar semi-major axis discussed in the present work is a refinement of *ab initio* models of lunar tides. This should make it possible to compare the theoretical predictions of the angular momentum lost by the Earth immediately with LLR results, without using the unreliable astrometric data at all.

Yet another promising way to exclude probable artefacts from the study of local Hubble expansion is usage of the artificial satellite systems, which do not suffer from the geophysical uncertainties discussed above (Dumin, 2001b). The suitable candidates may be space-based laser interferometers projected for searching the gravitational waves, such as *LISA* and *ASTROD*. In general, efficiency of such systems decreases sharply at low frequencies, beyond the operational range intended for the detection of gravitational waves; but, fortunately, the requirements on sensitivity for studying the local Hubble expansion are not so severe as in the case of gravitational waves. It is interesting to mention that measuring the long-term (secular) relativistic effects was not discussed till now by *LISA* team (e.g. Bender et al., 1998) but was declared as one of the main objectives (along with searching for gravitational waves) in *ASTROD* project (Bec-Borsenberger et al., 2000).

**CONCLUSIONS**

1. The lunar laser ranging combined with independent astrometric measurements of the Earth's rotation deceleration represents a valuable tool for searching the local Hubble expansion (Dumin, 2001a).

2. Preliminary analysis of the available data **under assumptions accepted in the present paper** gives the value of local Hubble constant about two times less than at intergalactic scales, which can be interpreted as influence of the uniformly-distributed dark matter (or the so-called "dark energy") on local dynamics of the space–time.

3. The above-mentioned result suffers from considerable geophysical uncertainties and can be verified by using artificial satellite systems, such as the projected space-based laser interferometers *LISA* and *ASTROD*.

4. If the effect of local Hubble expansion really exists, it will be evidently of great importance both for the cosmological studies and the theory of planetary evolution.

**ACKNOWLEDGMENTS**

I am grateful to Yu.V. Baryshev, P.L. Bender, A.D. Chernin, E.V. Derishev, J. Gaite, S.S. Gershtein, C. Hogan, C. Horellou, S.M. Kopeikin, V.N. Lukash, E.V. Mikheeva, S.M. Molodensky, W.-T. Ni, P.J.E. Peebles, N.I. Shakura, G. Tammann, M. Tinto, and A.V. Toporensky for valuable discussions and critical comments, as well as to D.P. Kirilova and V.D. Kuznetsov for various help in the course of work.

---

[4] Similarly, a secular increase in the martian orbit caused by the Hubble expansion will look like asteroidal perturbations rather than influence of variable *G*.

E-mail addresses of Yu.V. Dumin   dumin@yahoo.com, dumin@cityline.ru